\documentclass[a4paper,fleqn,usenatbib]{mnras}

\usepackage{newtxtext,newtxmath}

\usepackage[T1]{fontenc}
\usepackage{ae,aecompl}

\usepackage{mhchem}

\usepackage{graphicx}
\usepackage{subcaption}
\usepackage{float}
\usepackage{gensymb}
\usepackage{color}
\usepackage{booktabs,chemformula}
\usepackage[export]{adjustbox}

\usepackage{verbatim}
\usepackage{tabularx}

\usepackage{caption}
\usepackage{subcaption}
\usepackage{amsmath}

\usepackage{tikz}
\usepackage{hyperref}
\usepackage{longtable}
\usepackage{color}

\newcommand{\LamostGiants}{454,180}
\newcommand{\project}[1]{#1}
\newcommand{\lamost}{\project{LAMOST}}
\newcommand{\apogee}{\project{APOGEE}}

\newcommand{\tc}{\project{The Cannon}}

\newcommand{\teff}{T_{\rm eff}}
\newcommand{\logg}{\log_{10}(g/{\rm cm\,s}^{-2})}
\let\ACMmaketitle=\maketitle
\renewcommand{\maketitle}{\begingroup\let\footnote=\thanks \ACMmaketitle\endgroup}

\title[Mg--K stars in LAMOST]{On the discovery of K-enhanced and possibly Mg-depleted stars throughout the Milky Way\footnote{This paper includes data gathered with the 6.5 meter Magellan Telescopes located at Las Campanas Observatory, Chile.}}

\author[Kemp et al.]{Alex J. Kemp,$^{1}$\thanks{E-mail: ajkem1@student.monash.edu}
Andrew R. Casey,$^{1,2}$
Matthew T. Miles,$^{1}$
Brodie J. Norfolk,$^{1}$\newauthor
John C. Lattanzio,$^{1}$
Amanda I. Karakas,$^{1}$
Kevin C. Schlaufman,$^{3}$
Anna Y.~Q. Ho,$^{4}$\newauthor
Christopher A. Tout,$^{1,5}$
Melissa Ness,$^{6}$
Alexander P. Ji$^{7}$
\\
$^{1}$School of Physics \& Astronomy, Monash University, Clayton 3800, Victoria, Australia\\
$^{2}$Faculty of Information Technology, Monash University, Clayton 3800, Victoria, Australia\\
$^{3}$Department of Physics and Astronomy, Johns Hopkins University, Baltimore, MD 21218, USA\\
$^{4}$Cahill Center for Astrophysics, California Institute of Technology, MC 249-17, 1200 E California Blvd, Pasadena, Ca, 91125, USA\\
$^{5}$Institute of Astronomy, Madingley Road, Cambridge, CB3 0HA, United Kingdom\\
$^{6}$Department of Astronomy, Columbia University, 550 West 120th Street New York, New York 10027\\
$^{7}$The Observatories of the Carnegie Institution of Washington, 813 Santa Barbara St., Pasadena, CA 91101, USA
}

\date{Accepted 2018 XX XX. Received 2018 YY YY; in original form 2018 ZZ ZZ}

\pubyear{2018}

\begin{document}
\label{firstpage}
\pagerange{\pageref{firstpage}--\pageref{lastpage}}
\maketitle

\begin{abstract}
Stars with unusual elemental abundances offer clues about rare astrophysical events or nucleosynthetic pathways. Stars with significantly depleted magnesium and enhanced potassium ($[{\rm Mg}/{\rm Fe}] < -0.5$; $[{\rm K}/{\rm Fe}] > 1$) have to date only been found in the massive globular cluster NGC~2419 and, to a lesser extent, NGC~2808. The origin of this abundance signature remains unknown, as does the reason for its apparent exclusivity to these two globular clusters. Here we present 112 field stars, identified from \LamostGiants\ \lamost\ giants, that show significantly enhanced [K/Fe] and possibly depleted [Mg/Fe] abundance ratios.
Our sample spans a wide range of metallicities ($-1.5 < [{\rm Fe}/{\rm H}] < 0.3$), yet none show abundance ratios of [K/Fe] or [Mg/Fe] that are as extreme as those observed in NGC~2419. 
If confirmed, the identified sample of stars represents evidence that the nucleosynthetic process producing the anomalous abundances ratios of [K/Fe] and [Mg/Fe] probably occurs at a wide range of metallicities. This would suggest that pollution scenarios that are limited to early epochs (such as Population III supernovae) are an unlikely explanation, although they cannot be ruled out entirely. This sample is expected to help guide modelling attempts to explain the origin of the Mg--K abundance signature.
\end{abstract}




\section{Introduction}
\label{sec:intro}
NGC 2419 is the Milky Way's third most massive globular cluster, and its chemical composition makes it perhaps the most unusual star cluster in the Galaxy. Recent spectroscopic studies of red giant branch (RGB) stars in NGC 2419 revealed a strong anti-correlation between Mg and K abundances in nearly half of the studied stars, and weaker abundance relations in Si, Sc, Ca, Ti, and V with Mg \citep{mucciarelli2012,cohenkirby2012}.

A targeted search for [K/Fe] in other globular clusters (NGC 6752, NGC 6121, NGC 1904, NGC 104, NGC 6397, NGC 7099, and $\omega$ Centauri) and field stars concluded that all K abundance ratios fell within the bounds of the Mg-normal population in NGC 2419 \citep{carretta2013}, and no correlations between Mg and K were observed. Recent work by \cite{cerniauskas201747tuc} searched for unusual abundance patterns in Mg and K in NGC 104 by deriving abundances using 2dF/HERMES spectra, and a more comprehensive study involving over 400 stars in NGC 104, NGC 6752, and NGC 6809 by \cite{mucciarelli2017K} using FLAMES spectra \citep{cerniauskas201747tuc} both found no correlations between Mg and K, and no significant intrinsic spreads in either Mg or K.

NGC 2808 is the only cluster, other than NGC 2419, where an anticorrelation between Mg and K has been observed. All four of NGC 2808's known Mg-depleted stars show an anti-correlation with K \citep{mucciarelli2015}, although the amplitude of these abundance ratios is far weaker than those found in NGC 2419. The fact that the Mg-K anti-correlation is apparently confined to these two globular clusters implies either a small population of unusual polluter stars, or a single extremely massive polluter star. If it were the cumulative effect of many pollution events that was responsible for the signal, it would be expected that the anomalous abundances would be common among globular clusters.

Beyond the intrinsic scientific value of identifying the mechanism responsible for the puzzling Mg--K anti-correlation, a satisfactory explanation for the Mg--K signature may also offer insight into the underestimation of K abundances in the Milky Way predicted by Galactic chemical evolution models \citep{kobayashi2011}, or more broadly help to understand globular cluster formation and evolution. 

Here we use \lamost\ spectra to conduct the largest search to date for stars enhanced in potassium and depleted in magnesium. The discovery (or non-discovery) of such stars helps guide models that attempt to explain the Mg--K anti-correlation and related abundance phenomena. In particular, such a search provides strong evidence either for or against the uniqueness of the progenitor stars to globular clusters. In Section \ref{sec:method} we outline our methods to identify candidates enhanced in K and depleted in Mg, and the follow-up observations and abundance analysis for some of those candidates. In Section \ref{sec:discussion} we discuss our results and their implications for possible pollution mechanisms.

\section{Method \& Observations}
\label{sec:method}
\subsection{Candidate selection from \lamost\ spectra}
We used a set of \LamostGiants\ giant stars from the second \lamost\ data release \citep{luo2016vizier}. The spectra were placed at rest-frame on a common wavelength sampling from 3905\,\AA\ to 9000\,\AA\ and normalised by \citet{ho2017}. \tc\ \citep{ness2016,ho2017} was used to estimate effective temperature $\teff$, surface gravity $\logg$, metallicity [Fe/H], and mean $\alpha$-element abundance relative to iron [$\alpha$/Fe] using 9,952 stars in common between \lamost\ and the \apogee\ \citep{alam2015} survey. This process is referred to as 'label transfer'. Unless otherwise stated, these transferred labels are those referred to throughout this study. For details regarding the preparatory work, model generation, and label transfer between the \apogee\ and \lamost\ surveys we direct the reader to \citet{ho2017}. 

\begin{figure*}
\centering

\begin{subfigure}{1\textwidth}
\centering
\includegraphics[width=\textwidth=1]{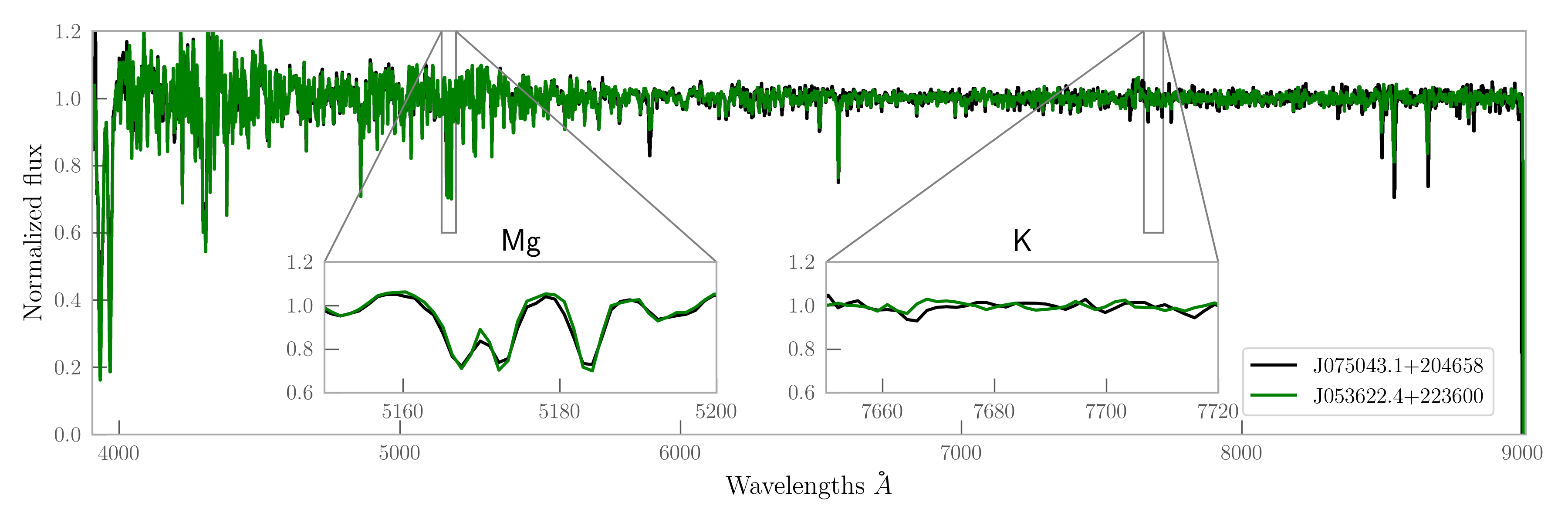}
\caption{Overlaid \lamost\ spectra for candidate J075043.1+204658 and J053622.4+223600, a star with similar stellar parameters}
\end{subfigure}

\begin{subfigure}{1\textwidth}
\centering
\includegraphics[width=\textwidth=1]{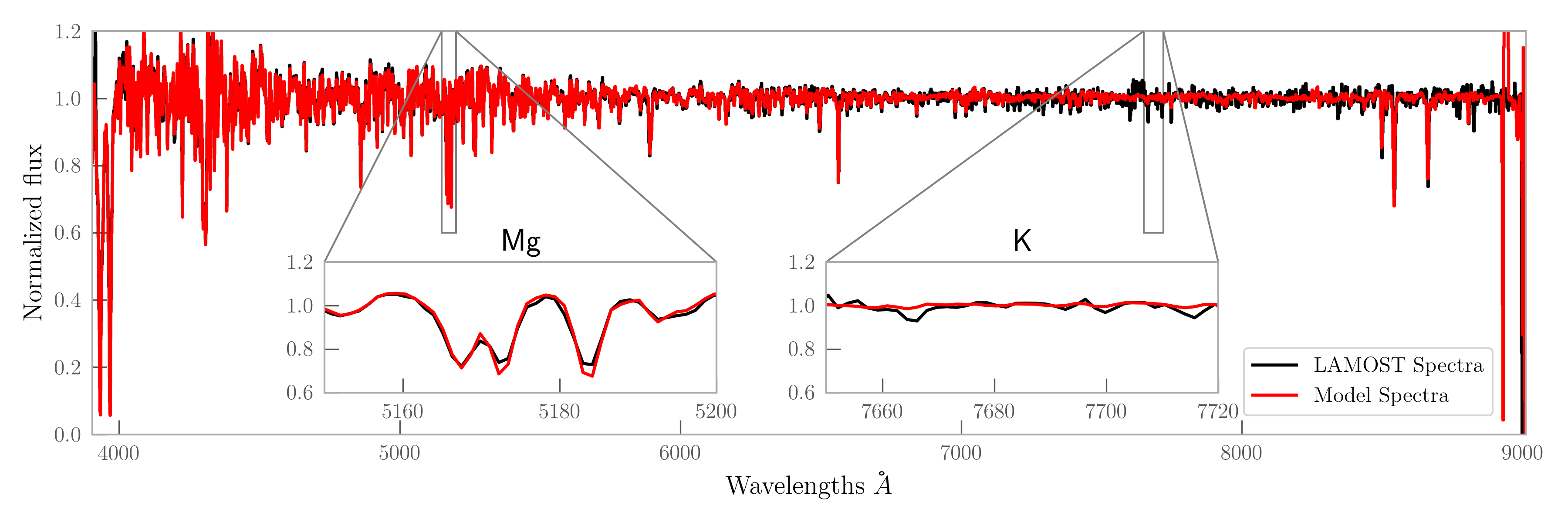}
\caption{Mg-depleted, K-enhanced candidate J075043.1+204658. $\teff$=4795, $\logg$=2.53, [Fe/H]=-0.47}
\end{subfigure}

\begin{subfigure}{1\textwidth}
\centering
\includegraphics[width=\textwidth=1]{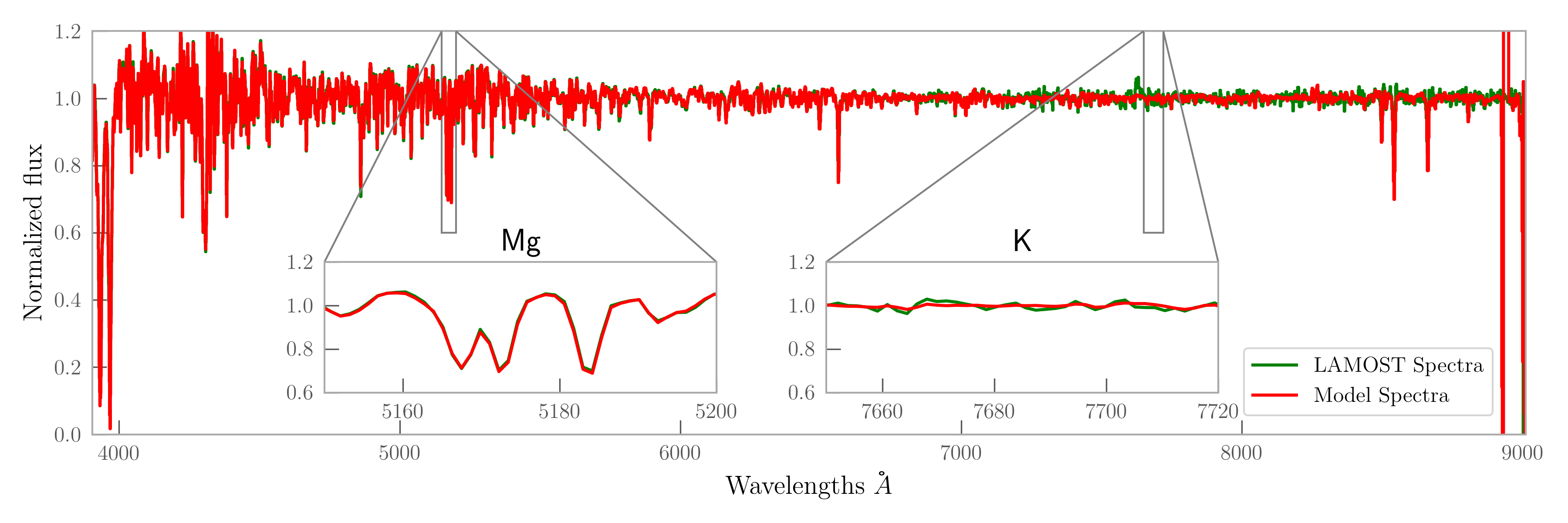}
\caption{Mg-normal, K-normal star J053622.4+223600. $\teff$=4804 K, $\logg$=2.53, [Fe/H]=0.062}
\end{subfigure}

\caption{Normalised \lamost\ spectrum for the Mg-depleted and K-enhanced candidate J075043.1+204658 and for the comparable star J053622.4+223600. J053622.4+223600 was selected by minimising the similarity norm defined as $n=(\frac{\Delta \text{T}_{\text{eff}}}{1000})^2 +\Delta \text{logg} ^2 + \Delta \text{[Fe/H]}^2$. The data are shown in black and the best-fitting data-driven model from \tc\ is in red. We zoom in around the magnesium triplet and potassium doublet, which we used to identify Mg--K candidates. Note that $[\alpha/{\rm Fe}]$ is a label in \tc\ model used and J075043.1+204658 shows Mg depletions relative to the estimated [$\alpha$/Fe] value.}
\label{posterchild}
\end{figure*}

We identified potential Mg-depleted and K-enhanced stars by searching for significant deviations in flux residuals. The flux residuals were calculated as the difference between the normalised \lamost\ flux and the best-fitting data-driven model flux from \tc\ ( $f_{\textrm{residual}} = f_{\textrm{data}} - f_{\textrm{model}}$). A positive residual implies a higher observed normalised flux than expected by the model (less stellar absorption than predicted by the model), while a negative residual implies a lower observed normalised flux than expected (more stellar absorption than predicted by the model). Figure \ref{posterchild} shows the \lamost\ spectrum and best-fitting model from \tc\ for J075043.1+204658, a Mg--K candidate star we identified, as well as the observed spectra for J053622.4+223600, the star with closest stellar parameters to our Mg--K candidate star, J075043.1+204658. 

We fit a Gaussian profile of amplitude $A$ to the flux residuals for all three absorption lines in the Mg triplet (5167\,\AA, 5172\,\AA, 5184\,\AA) as well as the K doublet (7665\,\AA, 7699\,\AA) for all \LamostGiants\ giants. For each star we recorded the amplitudes, widths (standard deviations) and wavelength (mean) about which the Gaussian profiles were fit. The amplitude was used as the measure of the discrepancy from the model spectra, and is analogous to the depth of the absorption line, and in principle linearly dependent on the equivalent width. This measure was favoured as more reliable than attempting to quantify line widths from the low resolution \lamost\ spectra.

We identified candidates by requiring that they match at least one of the following three quality filters:
\begin{enumerate}
\item We required the amplitude $A$ of the profile at the Mg 5184 \AA \ line to satisfy $A_{{\rm Mg} @ 5184} > 0.05$ and the amplitude of the profile at the K 7665\,\AA\ line to satisfy $A_{{\rm K}\,@\,7665} < -0.05$. Both amplitudes must also be measured at more than $3\sigma$ ($|A|/\sigma_{A} \geq 3$).
\item We required any two of the three Mg triplet lines to satisfy $A > 0.05$ and at least one K line to have $A < -0.05$, and for those amplitudes to have $|A|/\sigma_{A} \geq 3$.
\item We required the amplitude of any two of the three Mg triplet lines to have $A > 0$ and both K lines to have $A < 0$, and for the spectra to have a signal-to-noise ($S/N$) ratio of $S/N > 30$ in \lamost\ and a reported $\chi_{r}^2 < 3$ from \tc\ \citep{ness2016,ho2017}.
\end{enumerate} 
 
These filters identified 384 unique stars. We visually inspected every candidate (multiple times) and excluded stars that showed any evidence of being a false positive, including candidates that exhibited data reduction issues, apparent absorption that was narrower than the expected spectral resolution, as well as 75 stars that exhibited chromospheric emission at H$\alpha$, indicative of chromospheric activity consistent with being a young star that bas been misclassified as a giant. The distilled catalogue contains 112 candidate stars with spectra consistent with enhancements in K and depletions in Mg.

We note that although telluric absorption would usually swamp the K doublet lines, the contributions of telluric features have been removed from \lamost\ spectra as part of the data reduction process conducted by \lamost\ \citep{luo2016vizier}. Subsequent examination of the radial velocities for the (manually vetted) sample showed no behaviour indicating that the over-absorption at potassium wavelengths were due to telluric contamination.

We estimated [K/Fe] abundance ratios for all candidates by synthesising spectra to account for the flux residuals. We adopted the data-driven stellar parameters ($T_{\rm eff}$, $\log{g}$, [Fe/H]) and estimated microturbulent velocities $v_t$ as per Eq. 7 of \citet{kirby2008}. We used plane-parallel stellar photosphere models from \citet{marcs}, and compiled a list of atomic and molecular transitions from VALD \citep{vald}. We used Spectroscopy Made Easy \citep{sme} within the \texttt{iSpec} wrapper \citep{ispec} to calculate synthetic spectra.

It can be assumed that some stellar absorption is already accounted for by our data-driven model. In a sense, the data-driven model predicts a `typical' spectrum for a star with the given stellar parameters. This `typical` spectrum will include some contribution from all abundances (e.g., Mg, K, Al, Na), which is encapsulated by the overall metallicity [Fe/H]. That is to say that for a star of [Fe/H]$ = 0$, our data-driven model is predicting the spectrum of a \emph{typical} star at $[{\rm Fe/H}] = 0$, and that typical star would have $[{\rm K/Fe}] \sim 0$. We synthesise spectra to account for the flux residuals \emph{away} from the data-driven model.

We assume that deviations in flux around the potassium doublet are due to an increased [K/Fe] with respect to the model. We include a conservative 0.2\,dex systematic error floor applied in quadrature with the fitting errors to account for some component of the flux differences not being from the abundance of K, but instead due to noise in the \lamost\ spectra. We applied the same method (and error floor) to estimate [Na/Fe] from \lamost\ spectra. We note that while enhancements in [K/Fe] or [Na/Fe] can be estimated by synthesising the flux residuals, we cannot estimate [Mg/Fe] depletions in the same manner because it is difficult to separate the flux contribution of [Mg/Fe] from [$\alpha$/Fe] in the data-driven model. For this reason, we will adopt [$\alpha$/Fe] as a conservative upper limit for [Mg/Fe] such that $[{\rm Mg}/{\rm Fe}] \leq [\alpha/{\rm Fe}]$.\footnote{In some sense this is by construction because we required a flux depletion around the Mg lines to identify candidates, and our data-driven model includes the effects of [$\alpha$/Fe].}

\subsection{Follow-up observations with Magellan/MIKE}
We acquired high-resolution spectra of three of our candidates as ancillary targets to an existing observing campaign. The three targets (J075043.12+204658.0, J091825.49+172114.5, and J120032.60+024438.2) were selected based on their observability from Las Campanas Observatory, and we used the Magellan Inamori Kyocera Echelle (MIKE) spectrograph on the Magellan Clay telescope \citep{shectman2003magellan,bernstein2003mike}. Exposure times were calculated to achieve a S/N of 30 per pixel at 450\,nm, sufficient to verify the K-enriched and Mg-poor nature of these stars. We reduced the data using the \texttt{CarPy} package \citep{kelson2003}. We used smoothing spline functions to continuum-normalize individual echelle orders, before combining them to produce a continuum-normalized spectrum that is contiguous from 350\,nm to 950\,nm. With the stellar parameters from \lamost, we calculated [K/Fe] and [Mg/Fe] abundances from individual atomic lines by measuring equivalent widths with Gaussian absorption profiles \citep{castelli2004,sneden,casey2014}. From the Magellan/MIKE spectra we confirmed that the K lines in these candidates were unaffected by telluric contamination. The estimated abundances from Magellan/MIKE spectra are consistent (within the uncertainties) of our initial abundance determinations from \lamost\ (Figure \ref{KvsMg}), and are presented in Table \ref{data:magellan}. Additional atomic line abundances are included in Table \ref{data:atomiclines} in the Appendix.
\begin{table}
\centering
\caption{Summary of line abundances measured from Magellan/MIKE follow-up spectra.}
\label{data:magellan}
\begin{tabular}{lccccc}
\multicolumn{6}{c}{J075043.12+204658.0} \\
\hline
Species & $N$ & $\log_\epsilon(\textrm{X})$ & [X/H] & [X/Fe] & $\sigma/\sqrt{N}$ \\
\hline
\ion{Fe}{I} & 82 & 6.95 & $-$0.55 &    0.00 & 0.02 \\
\ion{Mg}{I} & 3  & 7.39 & $-$0.21 & $+$0.34 & 0.13 \\
\ion{K}{I}  & 2  & 5.33 & $+$0.30 & $+$0.85 & 0.07 \\
\hline
\\
\multicolumn{6}{c}{J091825.48+172114.5} \\
\hline
Species & $N$ & $\log_\epsilon(\textrm{X})$ & [X/H] & [X/Fe] & $\sigma/\sqrt{N}$ \\
\hline
\ion{Fe}{I} & 43 & 6.51 & $-$0.99 &    0.00 & 0.03 \\
\ion{Mg}{I} & 5  & 6.56 & $-$1.04 & $-$0.05 & 0.08 \\
\ion{K}{I}  & 2  & 5.32 & $+$0.29 & $+$1.28 & 0.31 \\
\hline
\\
\multicolumn{6}{c}{J120032.60+024438.2} \\
\hline
Species & $N$ & $\log_\epsilon(\textrm{X})$ & [X/H] & [X/Fe] & $\sigma/\sqrt{N}$ \\
\hline
\ion{Fe}{I} & 77 & 6.56 & $-$0.14 & 0.00 & 0.03 \\
\ion{Mg}{I} & 4  & 6.96 & $+$0.16 & $+$0.30 & 0.09 \\
\ion{K}{I}  & 2  & 4.96 & $+$0.72 & $+$0.86 & 0.11 \\
\hline
\end{tabular}
\end{table}

\begin{figure}
	\includegraphics[width=\columnwidth]{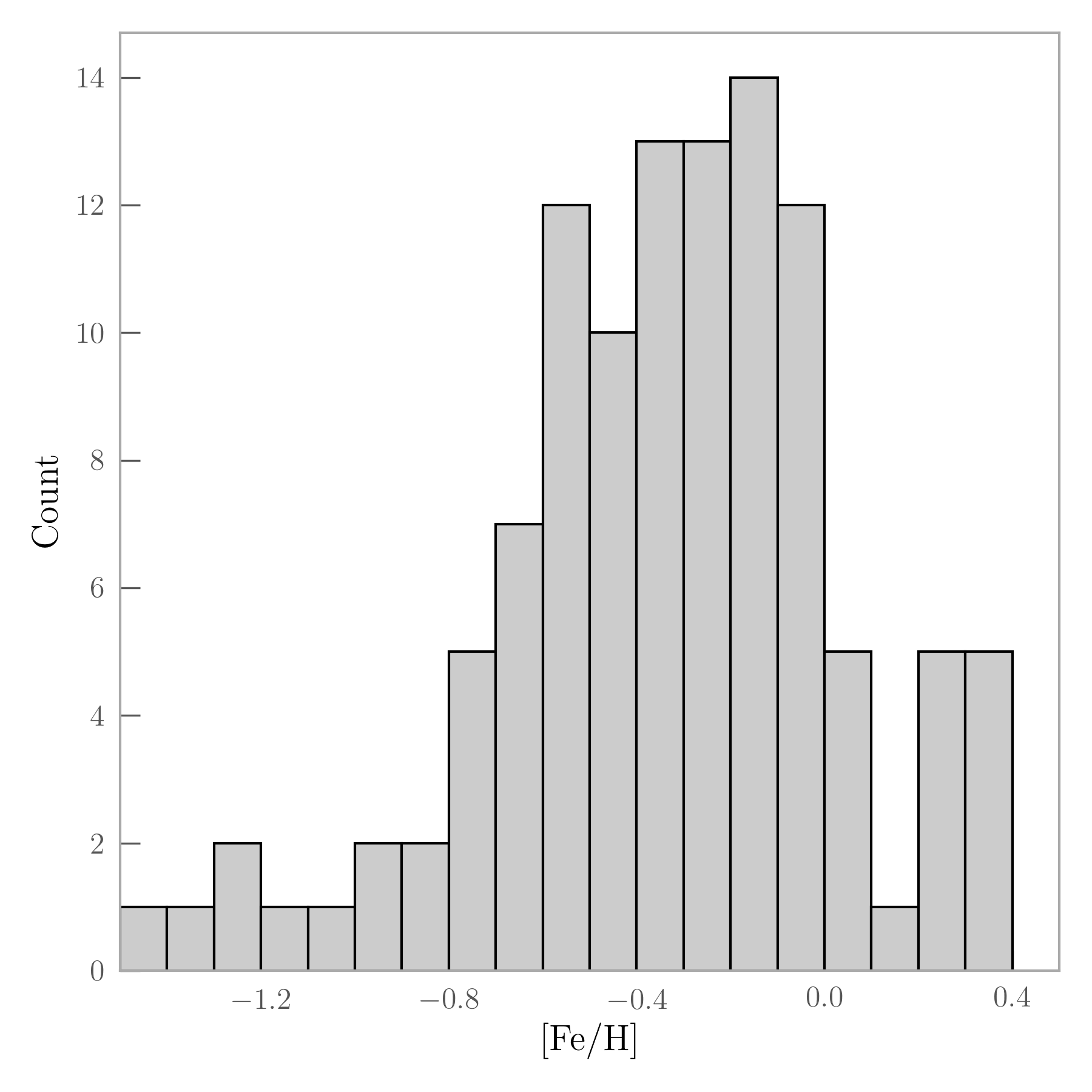}
    \caption{Metallicity distribution for all 112 candidate stars identified with K-enhancement and Mg-depletion. The stars apparently exist in significant numbers across a wide spread of metalicity.}
    \label{mhist}
\end{figure}

\section{Discussion}
\label{sec:discussion}

\subsection{Selection effects}
\label{sec:selectioneffects}

Our sample of \lamost\ giants carries with it certain selection effects from both the \apogee\ DR12 and \lamost\ samples. The scope of this the data-driven model is necesarily confined to the training sample, which is in turn confined in this case to stars common between \lamost\ and \apogee.\ For example, metal-poor stars with [Fe/H] $\lesssim-2$ are mostly absent from our sample because they were too dissimilar from the training stars for label transfer from \apogee. Similarly, the target selection process for \lamost\ is not easily invertible so we do not consider selection effects present in \lamost\ beyond comparing the candidate stars' metallicity distribution with that of \cite{ho2017}. We find that the frequency of Mg--K candidate stars is consistent (within the errors) of being flat with metallicity. 

\begin{figure}
	\includegraphics[width=\columnwidth]{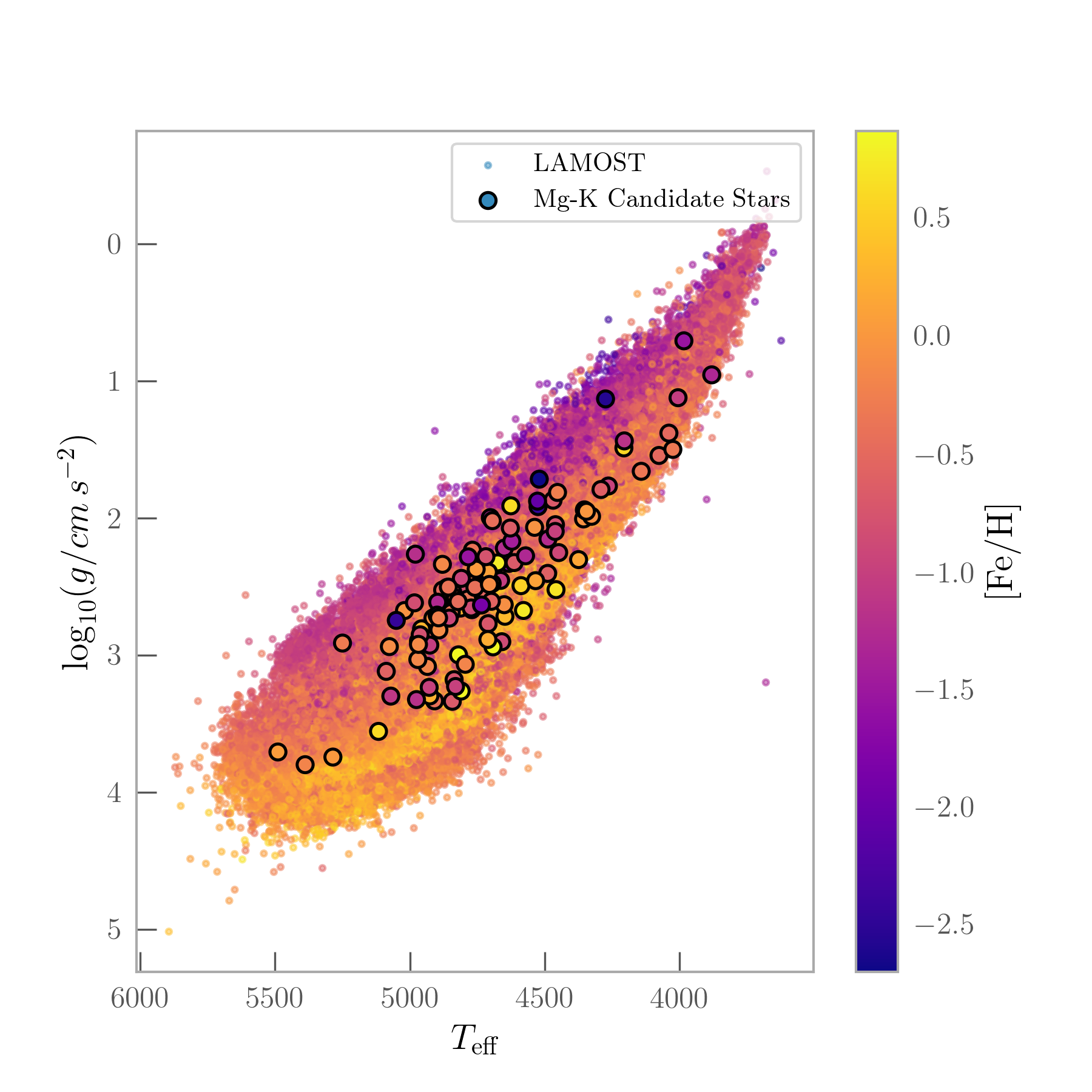}
    \caption{Effective temperature, surface gravity and [Fe/H] as per label transfer by the \tc\ for LAMOST, overlaid with the 112 candidates. The sample stars can clearly be seen to be spread out all along the giant branch.}
    \label{loggTeffFeh}
\end{figure}

\begin{figure}

\begin{subfigure}{0.5\textwidth}
\includegraphics[width=\columnwidth]{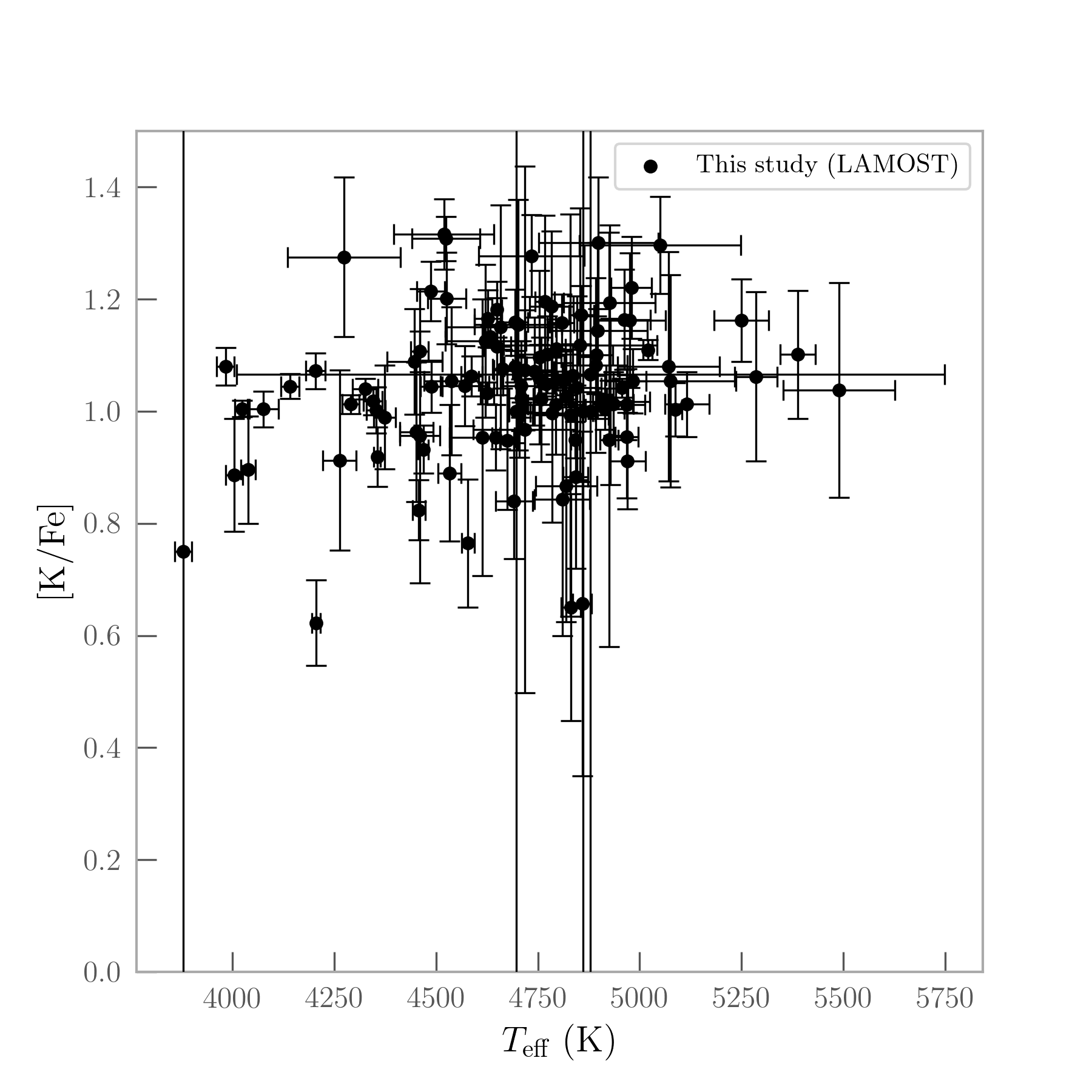}
\caption{[K/Fe] versus $\teff$}
\label{KvsTeff}
\end{subfigure}

\begin{subfigure}{0.5\textwidth}
\includegraphics[width=\columnwidth]{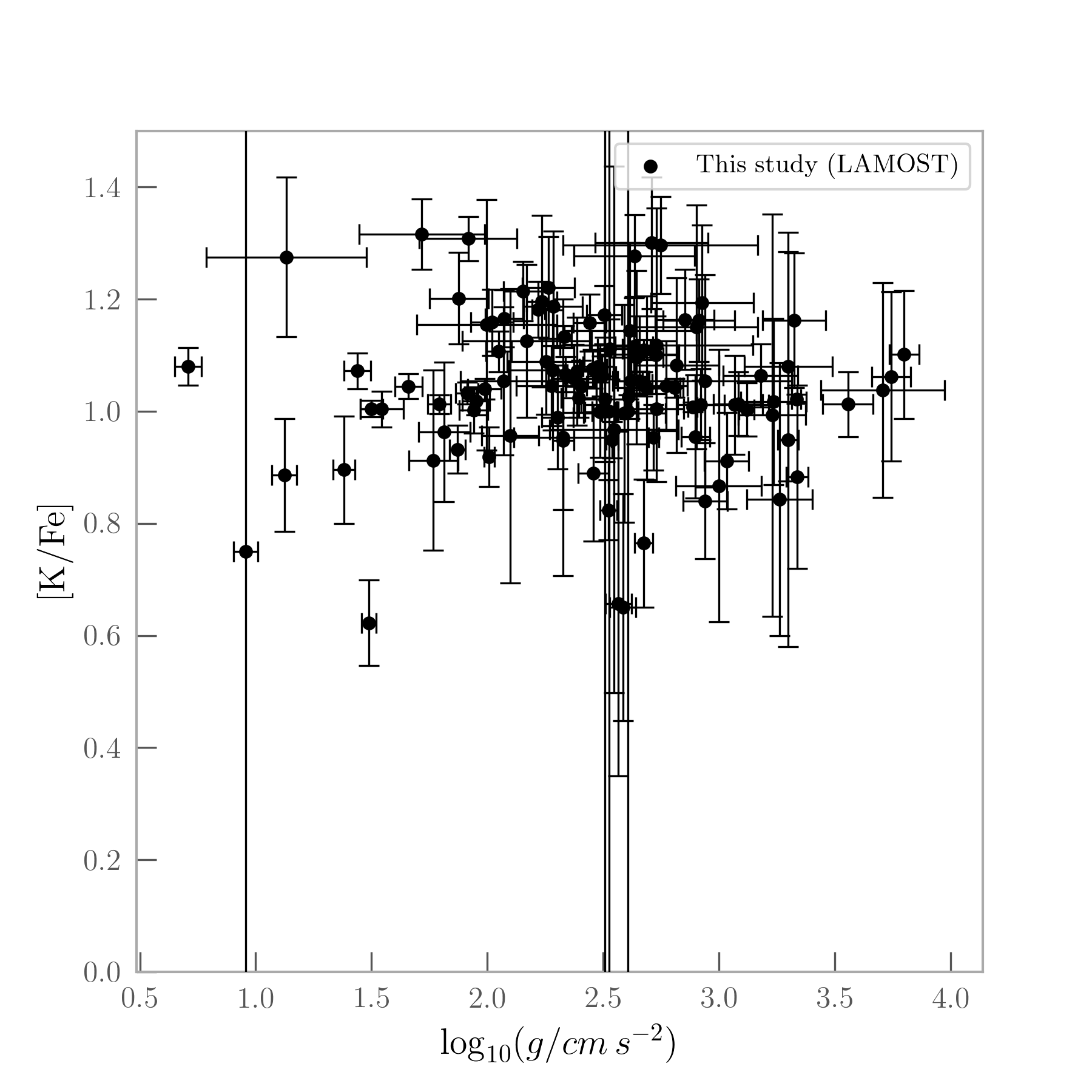}
\caption{[K/Fe] versus $\logg$}
\label{Kvslogg}
\end{subfigure}
\caption{No relationship is evident between [K/Fe] and either $\logg$ or $\teff$, as expected due to the use of the data-driven model.}
\end{figure}

Figure \ref{loggTeffFeh} shows that the identified candidates are distributed across the entire red giant branch, implying that the abundance signature is not associated with any single stage in evolution or any NLTE effects. One benefit of the data-driven approach is that NLTE effects need not be explicitly accounted for, as in a physically driven model. \tc\ predicts the flux of each star based its $\teff$, $\logg$, metallicity, and [$\alpha$/Fe] and then anomalously abundant stars are identified as those with flux significantly different from this prediction. Thus although it is possible the numeric value of the [K/Fe] abundances may be over-predicted due to NLTE effects, NLTE effects will not result in the identification of false positive candidates. This is consistent with what we observe: we find no relationship between the inferred [K/Fe] abundance and the stellar parameters $\teff$ or $\logg$ from \lamost\ (Figures \ref{KvsTeff} and \ref{Kvslogg}).

\subsection{Association with known globular clusters}
\label{sec:globclustasoc}
We cross-matched our candidates with the positions of known globular clusters \citep{harris1996} and found no obvious (or plausible) associations. The metallicity distribution function of our candidates (Figure \ref{mhist}) is consistent with the field, and the distribution reflects the original \LamostGiants\ sample of giants from which the candidates were selected.
No candidates have [Fe/H] $\approx -2$, the approximate metallicity of NGC 2419, nor are there many stars with [Fe/H] $\approx -1$, the metallicity of NGC 2808. Thus, we conclude that the candidates are not restricted to any specific metallicity, particularly not just to the metallicities of NGC 2419 and NGC 2808.

\subsection{Detailed chemical abundances}
In this section, we describe the abundance trends evident from the \lamost\ and Magellan/MIKE spectra and compare with stars identified as anomalous regarding their [Mg/Fe] and [K/Fe] abundances in other studies. Abundances as well as several physical and observational parameters are available for all candidate stars online, with a 10-star sample shown in Table \ref{data:lamost}.

\label{sec:abundances}

\subsubsection{[Na/Fe]}

\begin{figure}
\centering
\begin{subfigure}{0.35\textwidth}
\centering
	\includegraphics[width=\columnwidth]{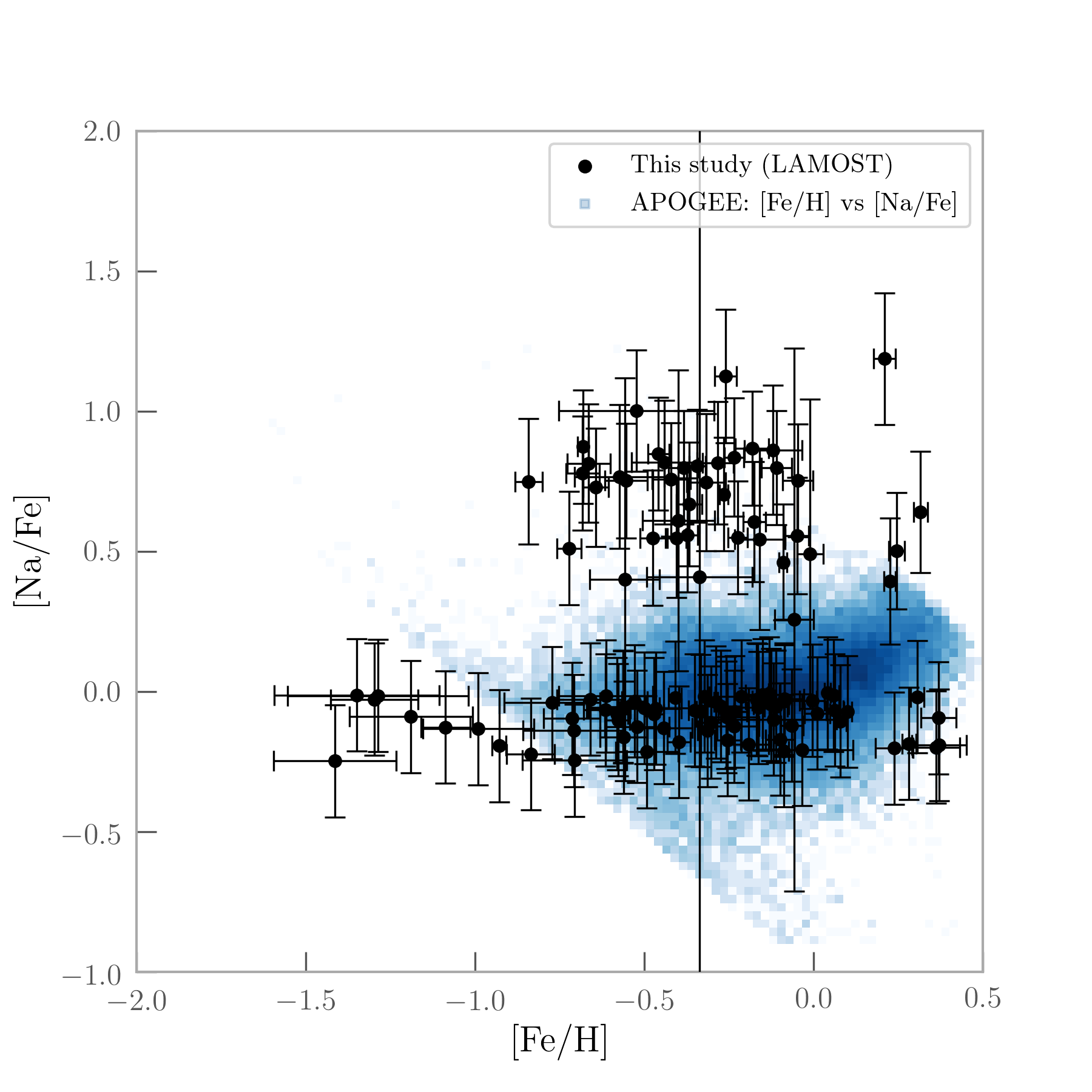}
    \caption{[Na/Fe] vs [Fe/H] for the candidate stars.}
    \label{NavsFeh}
\end{subfigure}

\begin{subfigure}{0.35\textwidth}
\centering
	\includegraphics[width=\columnwidth]{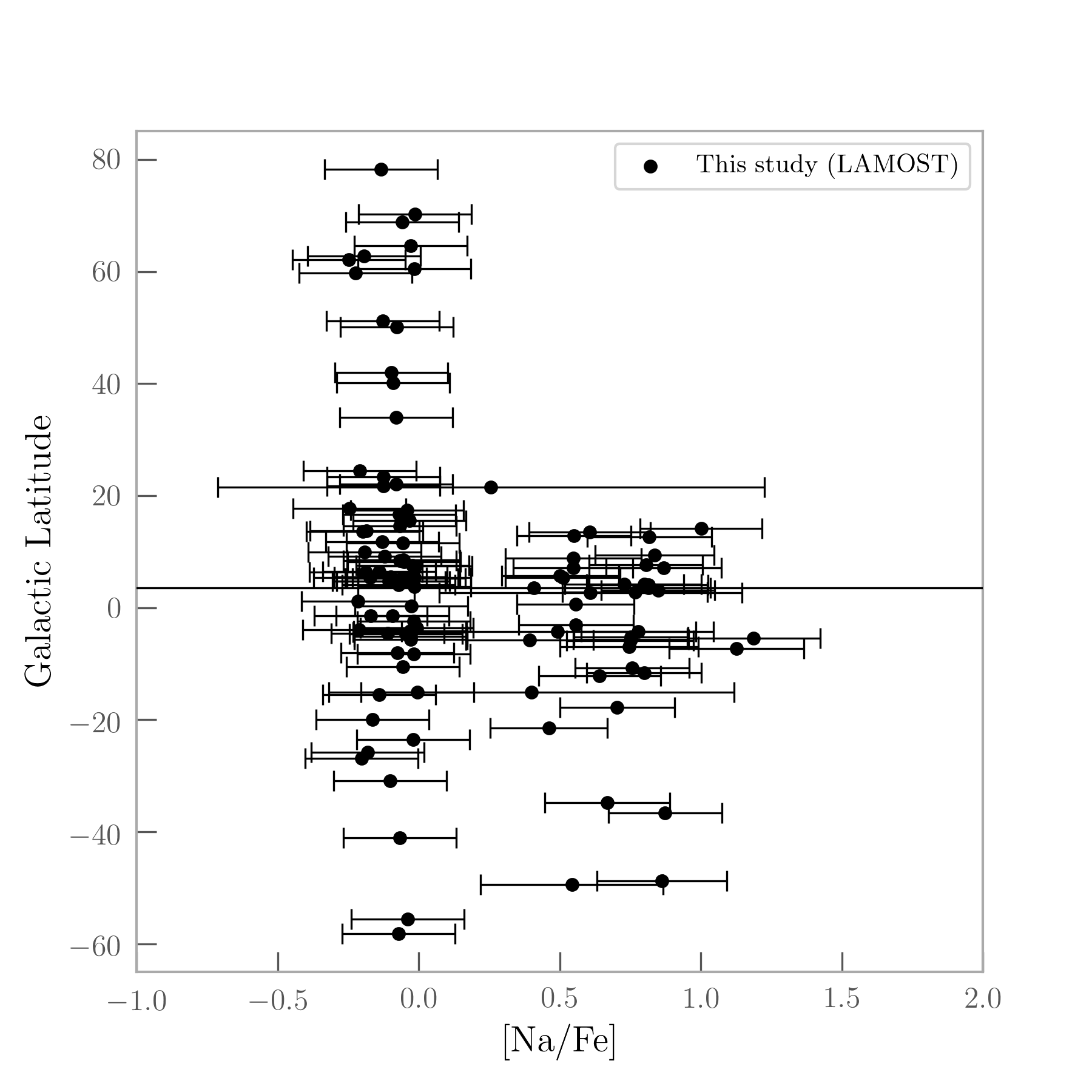}
    \caption{[Na/Fe] vs galactic latitude for the candidate stars.}
    \label{Navsb}
\end{subfigure}

\begin{subfigure}{0.35\textwidth}
\centering
	\includegraphics[width=\columnwidth]{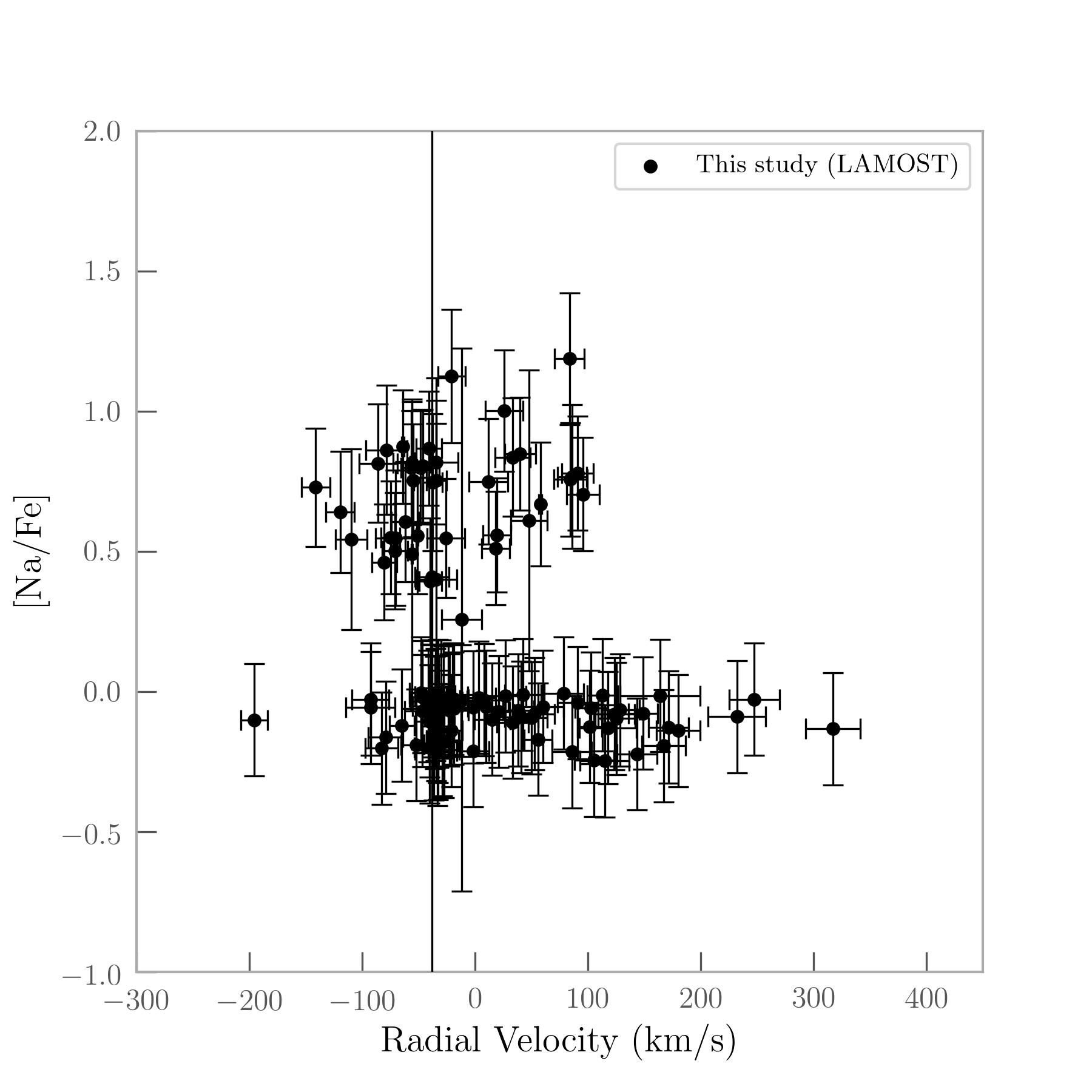}
    \caption{[Na/Fe] vs radial velocity for the candidate stars.}
    \label{Navsradvel}
\end{subfigure}

\caption{A strong bimodality in [Na/Fe] abundances of the candidate stars is observed, with a significant population displaying anomolously high abundances. Due to the apparent near-exclusivity of this population within the disk of the galaxy, and their notably higher E(B-V) values \citep{schlafly2011}, this phenomena is almost certainly due to interstellar dust.}
\end{figure}

The modelling work of \cite{prantzos2017} predicted an anti-correlation between Na and K which could result from a 180\,MK hydrogen burning environment, while observational data for NGC 2808 implies a positive correlation between Na and K. Upon examining the [Na/Fe] abundance ratios we estimated from \lamost\ spectra of our candidates, based on the doublet around 5850 \AA, we observed two distinct populations: one population making up approximately 30 percent of the sample with [Na/Fe] $> 0.5$, significantly higher than typical Galactic abundances of [Na/Fe] $\lesssim$ 0.3 \citep{kobayashi2011}, and a second population with [Na/Fe] abundance ratios consistent with typical Galactic levels. The two groups can be clearly distinguished in Figure \ref{NavsFeh}, where for comparison we also show the abundances of [Na/Fe] versus [Fe/H] from the \apogee\ survey.

However, we concluded that the high [Na/Fe] abundance ratios we derive are likely due to interstellar dust contributing. Nearly every star with apparently enriched [Na/Fe] showed notably higher E(B-V) \citep{schlafly2011}, and all but four stars with high [Na/Fe] are found in the disk of our Galaxy, at low absolute galactic latitudes and velocities (Figures \ref{Navsb} and \ref{Navsradvel}). Thus, we find no evidence for genuine  enhancement of [Na/Fe] among our Mg--K candidates.

\subsubsection{[Mg/Fe] and [K/Fe]}

\begin{figure}
	\includegraphics[width=\columnwidth]{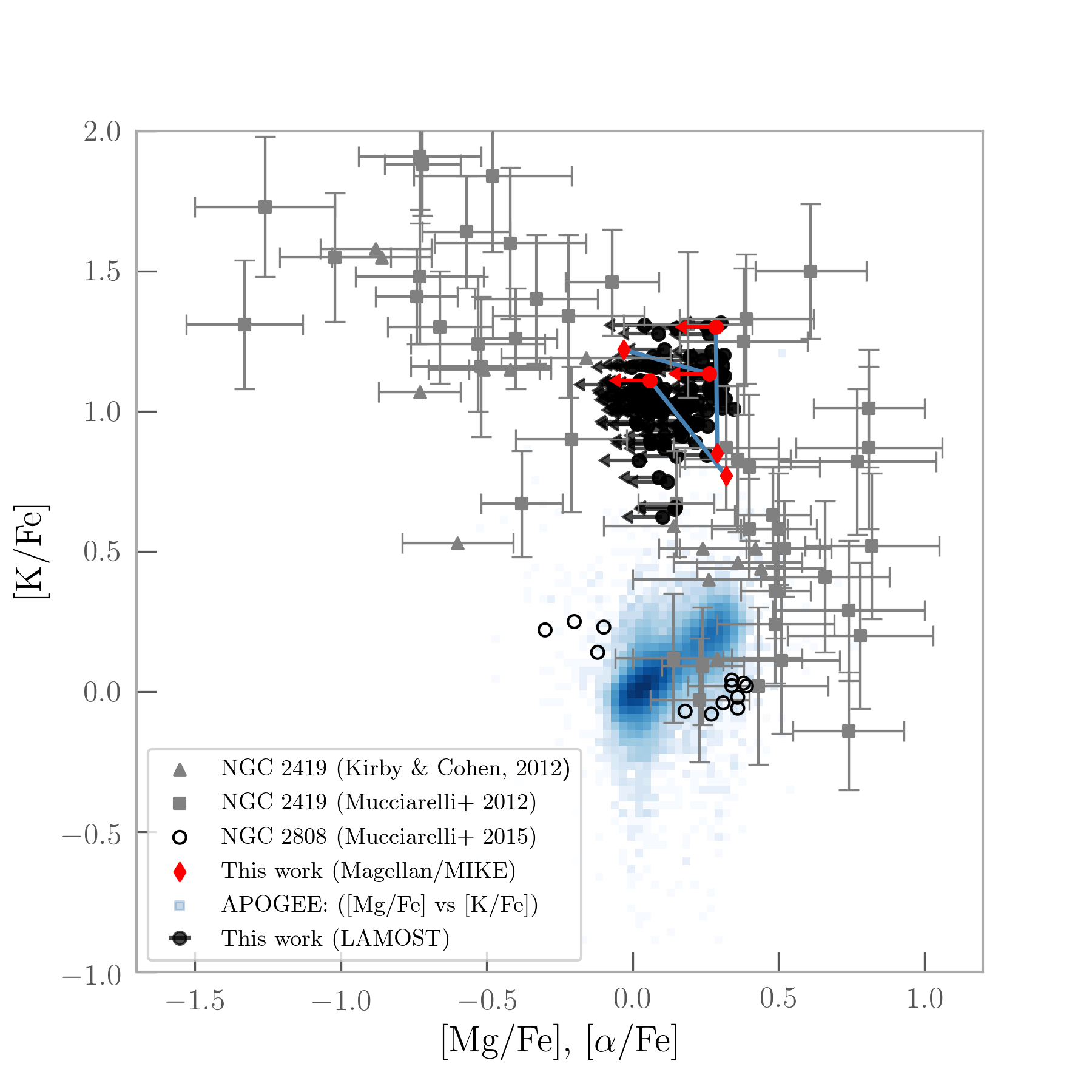}
    \caption{[K/Fe] vs [$\alpha$/Fe] for the candidate stars, overlaid with [K/Fe] vs [Mg/Fe] abundances for NGC 2419 and NGC 2808 \citep{cohenkirby2012, mucciarelli2012, mucciarelli2015} as well as the three candidates observed using Magellan/MIKE (red diamonds). These three stars' abundances as per the \lamost\ spectra are indicated in the figure (red circles). Note that in principle, [$\alpha$/Fe] acts only as an upper limit for [Mg/Fe] and is used as such for our sample in this figure due to difficulties in calculating reliable [Mg/Fe] from \lamost\ data.}
    \label{KvsMg}
\end{figure}

\begin{figure}
	\includegraphics[width=\columnwidth]{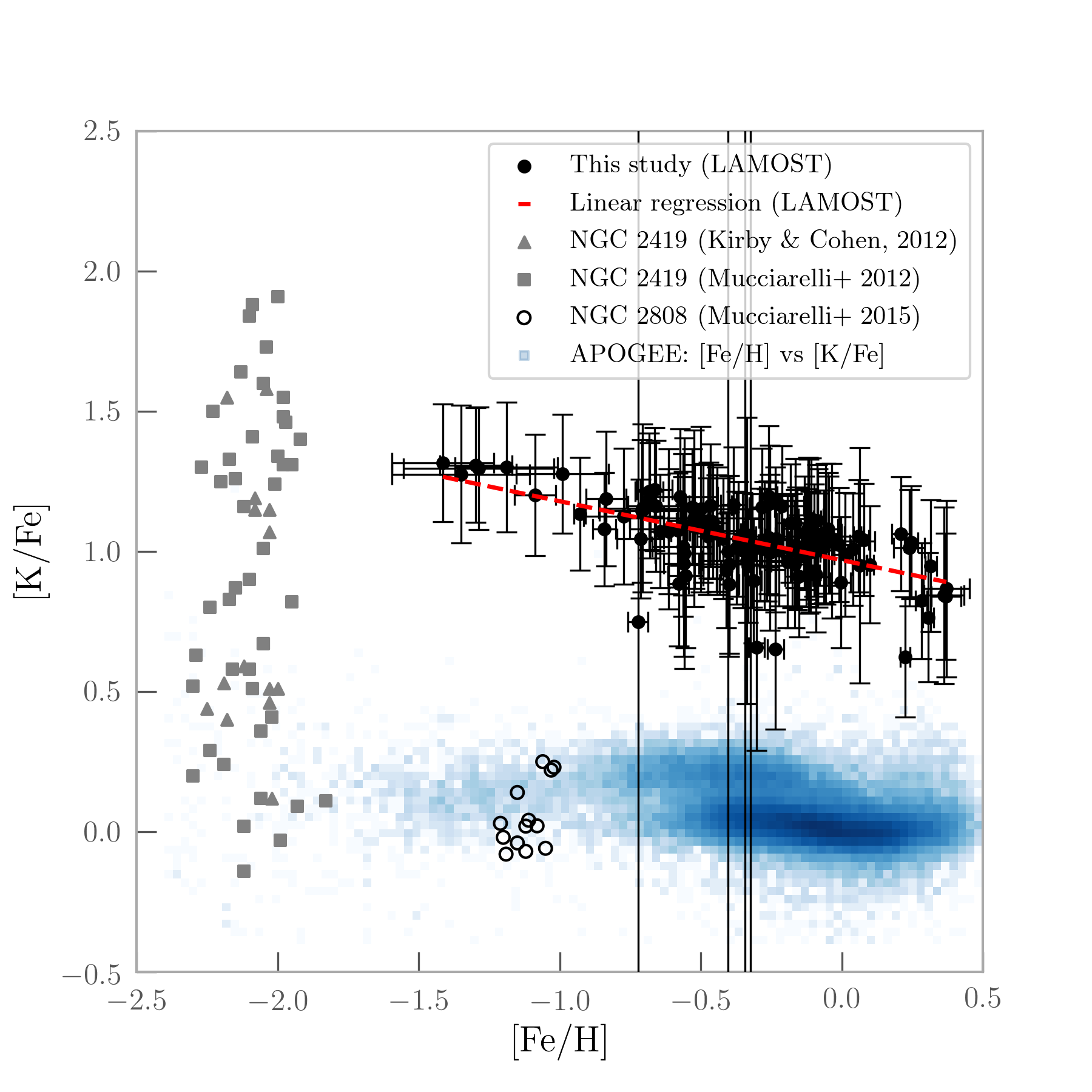}
    \caption{[K/Fe] vs [Fe/H] for the candidate stars, overlaid with [K/Fe] vs [Fe/H] for NGC 2419 and NGC 2808 \citep{cohenkirby2012, mucciarelli2012, mucciarelli2015}. The linear regression formula is \textrm{[K/Fe]$=-0.21 \times$[Fe/H]$+0.97$, with r$^2$=0.37}.}
    \label{KvsFe}
\end{figure}

Figure \ref{KvsMg} shows [K/Fe]  for the 112 star sample plotted against $[\alpha$/Fe] and overlaid with [K/Fe] from NGC 2419 and NGC 2808 plotted against [Mg/Fe] \citep{cohenkirby2012, mucciarelli2012, mucciarelli2015}. As described in Section \ref{sec:method}, we are forced to adopt [$\alpha$/Fe] as an upper limit for [Mg/Fe] for our sample stars as we are unable to estimate [Mg/Fe] from the LAMOST spectra. [K/Fe] vs [Mg/Fe] for the three stars from the sample observed using Magellan/MIKE and for over 60000 stars from APOGEE \citep{alam2015} is also shown. The \lamost\ abundances for the three stars observed by Magellan are also shown (red circles). We note that [K/Fe] for the candidate stars is generally lower than that of the Mg-depleted population in NGC 2419, but higher than that of the Mg-normal population. The [K/Fe] level for the candidates estimated by \lamost\ is significantly higher than the [K/Fe] abundances in NGC 2808.

Abundance ratios of [$\alpha$/Fe] in our sample fall within the lower limits for [Mg/Fe] of the Mg-normal population of both NGC 2419 and NGC 2808, although as we are unable to decouple [Mg/Fe] from $[\alpha$/Fe], the exact level of Mg depletion of the candidates is unclear. The [Mg/Fe] abundances derived from Magellan/MIKE's high resolution spectra are consistent with the Mg-normal population in NGC 2419. However, the \lamost\ spectral flux residuals with \tc\ model indicate an under-abundance of Mg relative to a 'typical' star in our sample of \lamost\ giants. The small sample of stars for which we have high resolution spectra, coupled with the conflict with the \lamost\ spectra, means we are unable to confidently make a statement regarding the level of [Mg/Fe] in our entire sample of 112 stars.

Given that the sample is almost certainly comprised of disk stars, it is relevant to also compare with typical Galactic abundances for [K/Fe]. In Figure \ref{KvsMg} and \ref{KvsFe} we make these comparisons with data from the \apogee\ survey \citep{alam2015}. Our sample's [K/Fe] appears definitively anomalous, with Galactic [K/Fe] around $0.3 \pm 0.2$\,dex, while the identified candidate stars have a spread in [K/Fe] from around 0.8 to 1.3 dex.

The [K/Fe] abundance ratios in our sample also show an anti-correlation with total metallicity, illustrated in Figure \ref{KvsFe}, which is not present when examining the bulk of Milky Way stars. It is interesting that the levels of [K/Fe] in the Mg-poor population of NGC 2419 appear consistent with this relation. 

For completeness, we caution that this trend may be an artefact of our analysis. [K/Fe] is calculated from the flux residuals ($f_{\textrm{residual}} = f_{\textrm{data}} - f_{\textrm{model}}$), and the data-driven model may already account for some level of [K/Fe] in the star due to the model representing the typical star at a given $\teff$, $\logg$, metalicity and [$\alpha$/Fe]. In principle, this has potential to cause spurious effects; however, it should also be emphasized that these effects would be within the uncertainties associated with estimating abundances from \lamost\ spectra. We recommend additional high-resolution observations to verify this trend.

\subsection{Impact on potential scenarios explaining the Mg--K anti-correlation}

The implications of the abundance ratios we find are intriguing. It is an astonishing fact that in such a large sample of \LamostGiants\ field stars, none were found that replicated (or exceeded) the extreme abundance pattern of NGC 2419 highlights the uniqueness of NGC 2419.

However, we repeat that the sample of \lamost\ giants  we searched had already been filtered of most of the stars that would have matched NGC 2419's metallicity. The 112 star candidate pool identified of K enhanced stars represents a unique collection of stars; while not as K enhanced as the extreme population in NGC 2419, they exhibit K abundances well above typical Galactic levels at a range of metallicities. It seems likely that whatever process is responsible for these anomalous field stars is similar to whatever caused the unusual stellar populations in NGC 2419 and NGC 2808.  Although a unifying explanation remains elusive, it is improbable that a processes acting to produce such high levels of [K/Fe] in the Milky Way is unrelated to whatever process is responsible for the extreme abundance ratios in NGC 2419. The apparent consistency of NGC 2419's extreme population with the anti-correlation between [K/Fe] and [Fe/H] observed in Figure \ref{KvsFe} further supports the idea that the process responsible for the stars in this study and the previously known anomalous stars in NGC 2419 are related.

If we allow a brief venture into the realm of speculation, the trend of [K/Fe] with iron may also provide a clue to the question of [Mg/Fe] abundances. While K is quite difficult to form, Mg is far more common. If we imagine that a cloud of gas formed with NGC 2419-like abundances of [Mg/Fe], [K/Fe], and [Fe/H] then we might expect that over time [Mg/Fe] would quickly approach typical galactic levels, as Mg is produced in large quantities and the cloud is enriched. However, the anomalously high level of [K/Fe] would vary only with increasing metallicity and therefore remain detectable in stars that formed from the cloud for far longer.

A recent attempt to replicate the Mg--K abundance signature in NGC 2419 by \cite{iliadis2016} also examined abundances of Si, Sc, Ca, Ti and V, elements reported as having weak correlations with Mg by \cite{cohenkirby2012}. The study aimed to constrain the temperatures and densities required to produce the chemical signature, and thereby provide insight into potential polluters.
Unless extremely high densities were invoked (i.e., $>\,10^8\,{\rm g\,cm}^{-3}$), it was found that temperatures between 100 and 200\,MK (depending on the density) were necessary to produce sufficient K to match the abundances observed. This constraint rules out many pollution sites, including core and shell burning of low-mass stars, high-mass and super-massive stars, and normal AGB stars. However, super-AGB (SAGB) stars were considered as potential candidates, with only a relatively small (roughly 10 to 20 MK) increase in temperature at the bottom of the envelope required to fall in the acceptable band of parameter space identified, implying that it is possible that the conditions for K production through the Ar-K reaction pathway could be met in some SAGB stars. The other potential candidates identified were novae, although the lack of detailed models of white dwarf accretion of metal-poor material adds considerable uncertainty. Based on current nova frequency in globular clusters determined by \cite{kato2013novae}, \cite{iliadis2016} conclude that the amount of material that would be produced by novae is at most 1 percent of the total required mass to pollute 30 percent of NGC 2419.

Another proposed  polluter candidate is a pair-instability super nova \citep[PISN;][]{carretta2013}. Unique to extremely massive Population III stars, these events involve the total destruction of the star, with no remnant left behind. The main argument for this idea is that the extreme rarity of these events, coupled with the huge masses of processed material released, could allow for the abundance signature in NGC 2419 to be the result of a single event, which would explain why a similar abundance trend is not seen in any other globular clusters. However, the signature odd-even proton number abundance pattern associated with PISNs, where odd Z elements are enhanced and even Z elements are depleted, appears absent from NGC 2419 \citep{carretta2013, cohenkirby2012}.

Our study presents a collection of stars in the disk of the Galaxy with a range of metallicities that show anomalously high levels of [K/Fe]. The spread in metallicities among our candidates is important. It implies that the process responsible for the Mg--K abundance signature may not be tied to a certain epoch in the Milky Way's evolution. These stars with enhanced potassium apparently tend to form at all ages, across all metallicities, so it is probable that the polluting source is extant at all metallicities. If we assume that the polluter star forms at a range of metallicities, then we can attempt to constrain scenarios that may be responsible for the Mg--K anti-correlation by disregarding polluters that are reliant on specific environments to form (e.g., low metallicity environments).

One class of candidates that can be immediately eliminated under such an assumption are Population III stars. A hyper-massive Population III star may be been a plausible explanation for NGC~2419 \citep{carretta2013} exclusively, particularly given its isolation from the Milky Way at a distance of $\approx 90\,{\rm kpc}$ and low metallicity. However, being unable to form throughout the life of the Galaxy makes it unlikely for Population III stars to be the polluters responsible for the stars we identify.

One possible polluter that \textit{could} form throughout the life of the Milky Way is a SAGB star with a binary companion. As emphasized by \cite{prantzos2017}, a significant challenge presented by the signatures of NGC 2419 and NGC 2808 is obtaining both a depletion in Mg and an enhancement in K. Reducing Mg is possible by converting \ce{^{24}Mg} to \ce{^{27}Al} through Mg--Al chains at temperatures around $75\,$MK, which is attainable in hot-bottom burning (HBB) environments in AGB stars. However, temperatures upwards of 150\,MK are required to produce K, and these temperatures might only be reached at the bottom of the convective envelope in SAGB stars \citep{iliadis2016}. This implies that the Mg depletion and K enrichment likely occurs at separate sites, assuming a relatively simple hydrogen burning system. 

We hypothesise that a SAGB-binary system may produce a [K/Fe] over-abundance in some stars within the Milky Way as follows. The SAGB star produces K \citep[destroying Na in the process;][]{prantzos2017}, that is then mixed throughout the star's envelope. The SAGB star then deposits its K-rich outer layers onto its lower-mass companion through accreting winds. The lower-mass companion continues to evolve long after the SAGB star has formed a white-dwarf remnant and, depending on its mass after accretion, either undergoes hot-bottom burning or, more likely, does not. If the companion star had sufficient mass to undergo hot-bottom burning, then Na could be produced \citep{prantzos2017} at the expense of Ne. Further, depending on the HBB temperature, the AGB star may also deplete Mg through Mg--Al chains. The material produced through HBB is mixed throughout the envelope including the surface of the star. In this scenario, it is the lower-mass binary companion stars that make up our sample of 112 giants. A direct test would be to determine whether the candidates are binaries with O-Ne white dwarf remnant companions, which would indicate that they formed from SAGB stars.

\section*{Acknowledgements}
We thank David W. Hogg (NYU), and Hans-Walter Rix (MPIA).
ARC is supported through an Australian Research Council Discovery Project under grant DP160100637.
AYQH is supported by a Fulbright grant through the German-American Fulbright Commission and a National Science Foundation Graduate Research Fellowship under Grant No. DGE-1144469. 
CAT thanks Churchill College for his fellowship and Monash University for hosting him as a Kevin Westfold distinguished visitor.
This research has made use of NASA's Astrophysics Data System.
Guoshoujing Telescope (the Large Sky Area Multi-Object Fiber Spectroscopic Telescope LAMOST) is a National Major Scientific Project built by the Chinese Academy of Sciences. Funding for the project has been provided by the National Development and Reform Commission. LAMOST is operated and managed by the National Astronomical Observatories, Chinese Academy of Sciences.

\bibliographystyle{mnras}
\bibliography{mgkbib}

\appendix

\section{Additional atomic line abundances from Magellan/MIKE}

\begin{table*}
\centering
\caption{Atomic line abundances measured from Magellan/MIKE follow-up spectra of three Mg/K candidates.}
\label{data:atomiclines}
\begin{tabular}{crrrcccccc}
\hline
& & & & \multicolumn{2}{c}{J075043.12+204658.0} & \multicolumn{2}{c}{J091825.48+172114.5} & \multicolumn{2}{c}{J120032.60+024438.2} \\

Species & $\lambda$ & $\chi$ & $\log{gf}$ & E.W. & $\log_\epsilon(\textrm{X})$ & E.W. & $\log_\epsilon(\textrm{X})$ & E.W. & $\log_\epsilon(\textrm{X})$ \\
& $({\textrm \AA})$ & $(\textrm{eV})$ & & $(\textrm{m\AA})$ & & $(\textrm{m\AA})$ & & $(\textrm{m\AA})$ \\
\hline
\ion{Mg}{1} & 4057.51 & 4.35 & $-$0.890 &        &       & 154.39 & 6.66 &        &      \\
\ion{Mg}{1} & 4167.27 & 4.35 & $-$0.710 &        &       & 123.88 & 6.27 &        &      \\
\ion{Mg}{1} & 4702.99 & 4.33 & $-$0.380 & 231.15 &  7.11 & 165.09 & 6.41 & 192.99 & 6.69 \\
\ion{Mg}{1} & 4730.04 & 4.34 & $-$2.389 &        &       &        &      &  76.41 & 7.18 \\
\ion{Mg}{1} & 5528.41 & 4.34 & $-$0.498 &        &       & 175.99 & 6.65 & 204.58 & 6.91 \\
\ion{Mg}{1} & 5711.09 & 4.34 & $-$1.724 & 129.32 &  7.38 &  85.64 & 6.80 & 115.01 & 7.09 \\
\ion{Mg}{1} & 6318.72 & 5.11 & $-$1.945 &  80.02 &  7.68 &        &      &        &      \\
\ion{K}{1}  & 7664.90 &  0.0 & $+$0.135 & 257.85 &  5.41 & 379.79 & 5.76 & 200.77 & 4.80 \\
\ion{K}{1}  & 7698.96 &  0.0 & $-$0.168 & 202.69 &  5.25 & 155.23 & 4.88 & 201.79 & 5.11 \\
\hline
\end{tabular}
\end{table*}

\begin{table*}
\centering
\caption{Tabulated data for 10 of the 112 candidate stars. Data for the full sample is available online.}
\label{data:lamost}
\begin{tabular}{cccccccccccccc}
\hline
\textbf{2MASSID} & \textbf{RA} & \textbf{DEC} & \textbf{S/N} & \textbf{V$_\textrm{r}$} & \textbf{T$_\textrm{eff}$} & $\boldsymbol \log{ \, \textbf{g}}$ & \textbf{{[}Fe/H{]}} & \textbf{{[}$\alpha$/Fe{]}} & \textbf{$\boldsymbol \chi_r ^\textbf{2}$} & \textbf{{[}K/Fe{]}} & \textbf{Error} & \textbf{{[}Na/Fe{]}} & \textbf{Error} \\ 
- & {[}deg{]} & {[}deg{]} & [pixel$^{-1}]$ & {[${\rm km\,s}^{-1}$]} & {[}K{]} & $[{\rm cm\,s}^{-2}]$ & {[}dex{]} & {[}dex{]} & - & {[}dex{]} & {[}dex{]} & {[}dex{]} & {[}dex{]} \\ \hline
J000908.89+124821.9 & 2.2871 & 12.8061 & 10 & -78.55 & 4892 & 2.82 & -0.12 & 0.33 & 0.19 & 1.08 & 0.26 & 0.86 & 0.23 \\
J002619.36+565612.1 & 6.5807 & 56.9367 & 31 & -39.57 & 4205 & 1.49 & 0.22 & 0.06 & 0.76 & 0.62 & 0.22 & 0.39 & 0.22 \\
J002907.61+354701.0 & 7.2817 & 35.7836 & 15 & -82.74 & 5116 & 3.56 & 0.24 & 0.17 & 0.18 & 1.01 & 0.21 & -0.20 & 0.20 \\
J003451.10+424543.0 & 8.7130 & 42.7620 & 15 & -79.15 & 4460 & 2.10 & -0.56 & 0.24 & 0.31 & 0.96 & 0.33 & -0.16 & 0.20 \\
J005649.98+391722.9 & 14.2083 & 39.2897 & 41 & -46.77 & 4578 & 2.67 & 0.31 & 0.1 & 0.44 & 0.76 & 0.23 & -0.02 & 0.20 \\
J010305.95+043445.9 & 15.7748 & 4.5794 & 90 & 20.69 & 4646 & 2.72 & 0.1 & 0.05 & 1.17 & 0.95 & 0.21 & -0.07 & 0.20 \\
J011949.36+063411.4 & 19.9557 & 6.5699 & 479 & -28.18 & 4820 & 2.61 & -0.29 & 0.09 & 2.38 & 1.03 & 0.20 & -0.04 & 0.20 \\
J013039.08+404843.8 & 22.6628 & 40.8122 & 71 & -80.34 & 4355 & 2.01 & -0.09 & 0.14 & 0.93 & 0.92 & 0.21 & 0.46 & 0.21 \\
J030101.42+560042.3 & 45.2560 & 56.0118 & 88 & -39.27 & 4878 & 2.53 & -0.32 & 0.09 & 0.67 & 1 & 8.70 & -0.02 & 0.20 \\
J032423.46+425429.6 & 51.0978 & 42.9082 & 23 & -56.74 & 4649 & 2.63 & -0.11 & -0.01 & 0.34 & 1.12 & 0.22 & 0.80 & 0.20 \\ \hline
\end{tabular}
\end{table*}

\bsp	
\label{lastpage}
\end{document}